# Working toward exposure thresholds for blast-induced traumatic brain injury: thoracic and acceleration mechanisms.

Michael W. Courtney, Ph.D., U.S. Air Force Academy, 2354 Fairchild Drive,
USAF Academy, CO, 80840 Michael.Courtney@usafa.edu

Amy C. Courtney, Ph.D., Force Protection Industries, Inc., 9801 Highway 78,
Ladson, SC 29456 amy_courtney@post.harvard.edu

**Abstract:** Research in blast-induced lung injury resulted in exposure thresholds that are useful in understanding and protecting humans from such injury. Because traumatic brain injury (TBI) due to blast exposure has become a prominent medical and military problem, similar thresholds should be identified that can put available research results in context and guide future research toward protecting warfighters as well as diagnosis and treatment. At least three mechanical mechanisms by which the blast wave may result in brain injury have been proposed – a thoracic mechanism, head acceleration and direct cranial transmission. These mechanisms need not be mutually exclusive. In this study, likely regions of interest for the first two mechanisms based on blast characteristics (positive pulse duration and peak effective overpressure) are developed using available data from blast experiments and related studies, including behind-armor blunt trauma and ballistic pressure wave studies. These related studies are appropriate to include because blast-like pressure waves are produced that result in neurological effects like those caused by blast. Results suggest that injury thresholds for each mechanism are dependent on blast conditions, and that under some conditions, more than one mechanism may contribute. There is a subset of blast conditions likely to result in TBI due to head acceleration and/or a thoracic mechanism without concomitant lung injury. These results can be used to guide experimental designs and compare additional data as they become available. Additional data are needed before actual probabilities or severity of TBI for a given exposure can be described.

## Introduction

Primary blast-induced TBI remains a difficult injury to predict and diagnose (Hoge et al., 2008). Therefore, effective preventive measures and treatments remain elusive. Primary blast-induced TBI is not new (Jones et al., 2007), but it has increased in documented incidence (and public prominence) in modern conflicts as more warfighters are exposed to nonlethal levels of blast and as body armor is more effective in preventing lethal penetrating injuries (Cernak et al., 1999a; Moore et al., 2008).

In major 20th century conflicts, blast lung was a prominent injury, and scientists worked to establish injury thresholds based on exposure (for examples, Yelverton et al., 1973; Stuhmiller, 1994). The Bowen curves (Bowen et al., 1968) were based on a large body of research and are considered a standard to guide military training, planning and equipment design. Despite the research effort and useful findings, no lung injury threshold has proven widely applicable, and improvements as well as alternate approaches for complex blast conditions have been suggested (Bass et al., 2008; Axelsson and Yelverton, 1996; Stuhmiller et al., 1998).

With a few notable exceptions (Cernak et al., 1999b), research in blast-induced TBI has focused on diagnosis, often months after the initial injury (Trudeau et al., 1998; Schwab et al., 2007; Scott et al., 2005), and on effects of exposure to pressure waves from shock tubes or actual blast exposures in animal models (Irwin et al., 1999; Cernak et al., 2001a, 2001b; Yang et al., 1996). Animal studies have clarified the nature of the neural injury and suggested mechanical mechanisms by which the blast wave may result in brain injury. However, experimentally isolating a particular mechanism is challenging. It is also difficult to compare studies or to place experimental conditions or results in the context of human exposure to real-life blast conditions.

Exposure thresholds for brain injury due to the mechanical mechanisms of blast wave transmission are desired. Available data can be used to identify regions of interest for these thresholds. Rather than provide definitive predictive ability, regions of interest provide organizational structures for interpreting existing data and designing future studies.

**Materials and Methods**

*Three candidate mechanical mechanisms for primary blast-induced TBI*

Pressure waves can injure neural cells (Suneson et al., 1989; Saljo et al., 2003; Chen et al., 2009), but how blast waves result in brain injury is a topic of ongoing research. Three candidate mechanical mechanisms for primary blast-induced TBI have emerged:[1]

1. A thoracic mechanism, as described by Cernak et al. (Cernak, 2005; Battacharjee, 2008), by which a blast pressure wave enters the thorax and leads to brain injury. A recent review discusses evidence for this mechanism (Courtney and Courtney, 2009).
2. Head acceleration – translational and rotational – is known to be a mechanism for TBI (Zhang et al., 2004; Krave et al., 2005). FEM and numerical modeling of head models exposed to blast waves demonstrates the plausibility of primary acceleration-induced TBI (Stuhmiller et al., 1998) in which the blast wave itself induces injurious head acceleration (separately from the potential for secondary and tertiary blunt force head trauma (Finkel, 2006)).
3. Direct cranial entry of blast waves. For example, blast waves pass through the (thin) cranium of rats almost unchanged (Chavko et al., 2007) and through the thicker cranium of pigs maintaining about two-thirds magnitude (Bauman et al., 2009); however, it is still under investigation how a blast wave interacts with the human cranium (Singer, 2008). Finite Element Modeling (FEM) of blast waves applied to human head models illustrates the potential for blast waves to enter the cranium directly (Taylor and Ford, 2009; Nyein et al., 2008; Stuhmiller et al., 1998), a mechanism which may include skull flexure (Moss et al., 2009).

These proposed mechanisms need not be mutually exclusive, and important implications are discussed later in this paper. Also, it is unclear what features of the blast wave correlate most highly with neural injury. The present study aims to develop proposed injury thresholds based on peak effective pressure and positive pulse duration for thoracic and acceleration mechanisms. Due to space limitations and emerging studies, development of proposed thresholds for a direct cranial mechanism is not addressed.

*Thresholds for blast-induced lung injury*

Bowen et al. (1968) analyzed blast experiments involving a large total sample of 2097 subjects from thirteen mammalian species ranging in mass from 0.02 kg to 200 kg. Positive pulse duration of experimental blasts ranged from 0.24 to 400 ms. Body mass was a sufficient scaling parameter for the observed responses, and they developed what are now called the Bowen curves based on a body mass of 70 kg.

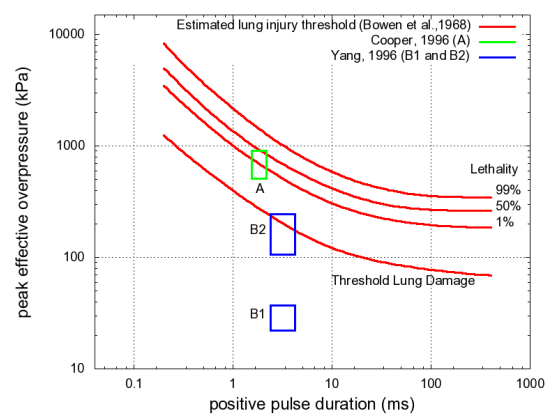

**Figure 1.** *Blast lung thresholds (Bowen et al., 1968) and later blast experimental data showing the experimental range of pressure wave characteristics in which respiratory tract injury was observed. A: 16-60 kg pigs, (Cooper, 1996). B1: threshold and B2: severe lung injury in 15-42 kg sheep (Yang et al., 1996) with 25% uncertainty shown in the peak effective overpressure.*

The Bowen curves are a frequently referenced standard. However, no lung injury threshold has proven applicable in a wide variety of blast situations and improvements have been suggested (Bass et al., 2008). The Bowen curves referenced herein apply for free-field blasts in which the long axis of the subject is perpendicular to the direction of travel of the blast wave. Chest wall velocity (Axelsson and Yelverton, 1996) and normalized work (Stuhmiller et al., 1998) have been used to address complex blast wave exposures in enclosed spaces.

---
[1] Toxic fume inhalation associated with blast exposure is an example of a non-mechanical TBI mechanism (Born, 2005).

For a given positive pulse duration, injury thresholds (peak pressures) can be determined. Note that different sources of blast vary in positive pulse duration (Table 1).

*Table 1 Typical positive pulse durations for representative sources of injurious blast waves.*

| Blast Type | Positive Pulse Duration (ms) |
|---|---|
| Underwater blast (Parvin et al, 2007) | 0.01 – 0.2 |
| IED/mine (Alley 2009) | 0.25 – 0.7 |
| Larger conventional high explosives (Bowen et al., 1968) | 1 - 10 |
| Fuel-air/nuclear (Liu et al., 2008) | 20 - 100 |

Results of two blast studies are plotted with the Bowen curves in Figure 1 (Cooper, 1996; Yang et al., 1996). Both involved animal subjects exposed to free-field blasts but under different conditions. Yang et al. (1996) observed lower damage thresholds than expected based on the Bowen curves. This may be because smaller animals were tested, and/or because more sensitive damage detection criteria were used. Cooper used lung edema (weight of exposed lungs divided by weight of control lungs) as the injury criterion. Yang et al. took tissue samples after exposure and examined them using light and electron microscopy. The threshold pressure value was taken as the lowest pressure at which the slightest injury was observed in the lung tissue.

*Thresholds should be identified to inform blast-induced TBI research*

Data are not available to develop exposure thresholds for blast-induced TBI with definitive predictive ability, and there are a number of challenges to developing general relationships between blast conditions and injury. For the development of the Bowen curves, scaling based on body weight was used to develop lung injury thresholds for humans. In blast-induced TBI, scalings are probably dependent on the mechanism of transfer of the blast wave and have not been determined with certainty for each mechanism. For example, a vagally-mediated reflex might scale by lung size or body mass. Depending on the details, several scalings are possible for the transmission of energy from the thorax to the brain via pressure waves. Blast-induced TBI due to head acceleration mechanisms might scale by brain size (Gibson, 2006). Susceptibility to TBI due to direct cranial mechanisms might scale according to skull size and/or thickness. In this paper, no scalings were used.

*Using available data to estimate blast-induced TBI thresholds*

A thoracic mechanism of primary blast-induced TBI has been supported by recent studies (Cernak et al., 1996; Knudsen and Oen, 2003; Gryth, 2007; Krajsa, 2009). Ideal data that would support definitive thresholds are unavailable and might still be several years away. Yet useful data are available and it is prudent to analyze what those data suggest. TBI-related results of few nonlethal blast experiments are available in the open literature. However, data from the related research areas of behind-armor blunt trauma (BABT) and ballistic pressure waves (BPW) are also instructive.

Cripps and Cooper (1996) showed that the coupling of blast waves into the thoracic cavity can be understood in terms of the chest wall acceleration imparting shock waves to the underlying tissue. They argued that the coupling of external pressure waves to the body necessitates gross and/or minute motion of the body wall. Therefore, insults producing similar motions of the body wall initiate pressure waves (compressive and/or shear) in the body. Such pressure waves can be initiated by the rapid deformation of body armor due to ballistic impact (Cannon, 2001; Roberts et al., 2007; Merkle et al., 2008).

A ballistic pressure wave (BPW) is generated when a ballistic projectile enters a viscous medium (Lee et al., 1997), such as the body. As the projectile loses energy over a short distance, large forces are generated that create pressure waves (force per unit area) that propagate through the medium.

In both BABT and BPW, pressure waves with blast-like characteristics are propagated through the body (Liu et al., 1996; Cannon, 2001; Courtney and Courtney, 2007). Experimental and theoretical studies (Harvey and McMillen, 1947; Lee et al., 1997; Courtney and Courtney, unpublished data) have shown that ballistic pressure waves produce a steep shock front followed by a near-exponential decay of pressure. Neurological symptoms and damage have been reported that are similar to those observed in blast-induced TBI (Axelsson et al., 2000; Cernak et al., 2001; Drobin et al., 2007; Suneson et al., 1987, 1988, 1990a,b; Wang et al., 2004; Krajsa, 2009).

*Thoracic mechanism of blast wave transmission leading to primary blast-induced TBI:*
Several BABT studies report data useful for development of exposure thresholds for blast-induced TBI via a thoracic mechanism (Gryth et al., 2007; Drobin et al., 2007; Gryth et al., 2008, Gryth 2007). In an experiment by Gryth et al. (2007), the amount of energy transferred to the thoracic tissues of large swine (60 kg) in two experimental groups was approximately 500 J and 760 J (Courtney and Courtney, 2009). The EEG (electroencephalogram) signals of two of ten subjects in the 760 J group flatlined immediately, and five subjects died before the end of the two-hour observation period. Two of eight test subjects in the 500 J group also died within two hours. Surviving subjects in both groups experienced EEG suppression that usually resolved within a few minutes.

In another test of seven swine, 320 J of energy was transferred to the thoracic tissues (Drobin et al., 2007; Courtney and Courtney, 2009); all subjects survived, but in five a reduction in EEG activity occurred after the shot. In each case, EEG activity returned to baseline within two minutes. These EEG changes are similar to those observed by Goransson et al. (1988) after penetrating injury of the hind limbs of swine. Observation of significant EEG suppression is the criteria for including this data as indicative of remote effects on the brain due to chest wall acceleration produced by bullet impact.

*Table 2: Data illustrating the process of converting impact energy (3125 J) and armor back face deformation (d) from BABT studies to peak effective overpressure for comparison with blast studies. The mechanical work (W) in tissues is computed from the formula in Appendix A of Courtney and Courtney (2009). The effective force ($F_{eff}$) is W/d, where d has been converted to meters, so the unit of force is newtons. $M_{eff}$ is the effective mass of the chest wall (Axelsson and Yelverton, 1996). The effective chest wall acceleration is computed from Newton's second law, $a_{eff} = F_{eff}/M_{eff}$. To compute effective peak overpressure from chest wall acceleration, data from Cooper (1996), who measured peak chest wall acceleration vs. peak overpressure in a series of blast experiments in sheep with accelerometers attached to the ribs, were used to construct a mathematical model (via non-linear least squares regression ensuring zero acceleration at zero pressure) of $P_{eff}$ vs. $a_{eff}$. The effective interaction time, $t_{eff}$, is computed from kinematic considerations, assuming the average bullet velocity during impact is 1/3 of the impact velocity.*

| d mm | W J | $F_{eff}$ N | $M_{eff}$ kg | $a_{eff}$ m/s$^2$ | $P_{eff}$ kPa | $t_{eff}$ ms | Note |
|---|---|---|---|---|---|---|---|
| 42 | 860 | 20477 | 2.0 | 10239 | 441 | 0.158 | i |
| 40 | 772 | 19297 | 2.0 | 9648 | 433 | 0.150 | ii |
| 34 | 524 | 15411 | 2.0 | 7705 | 402 | 0.128 | ii |
| 28 | 316 | 11287 | 2.0 | 5643 | 360 | 0.105 | iii |

i. Gryth, 2007
ii. Gryth et al., 2007
iii. Drobin et al., 2008

Table 2 shows the results of relating cerebral effects observed in BABT studies to equivalent blast parameters. Newton's second law was applied using the peak retarding force (Courtney and Courtney, 2009) and chest wall effective mass (Axelsson and Yelverton, 1996) to estimate peak chest wall acceleration (Table 2). Then data from Cooper (1996) was used to estimate an equivalent peak effective overpressure to compare BABT with blast injury results. A simpler model of estimating peak effective pressure as peak retarding force divided by effective area (Axelsson and Yelverton, 1996) yields peak pressures from 1.8 to 2.6 times smaller than the non-linear model built from the measured peak acceleration vs. peak pressure data in Cooper (1996). Both Eqn. 1 from Axelsson and Yelverton (1996) and the experimental data from Cooper (1996) support a non-linear relationship between chest wall acceleration and effective overpressure.

Additional advantages of applying this BABT data to blast-induced TBI are that human-sized subjects were used, the insult was quantifiable, and the effects were carefully measured. However, the analysis employed for comparison with results of blast studies introduces uncertainty, thereby limiting the precision of the results.

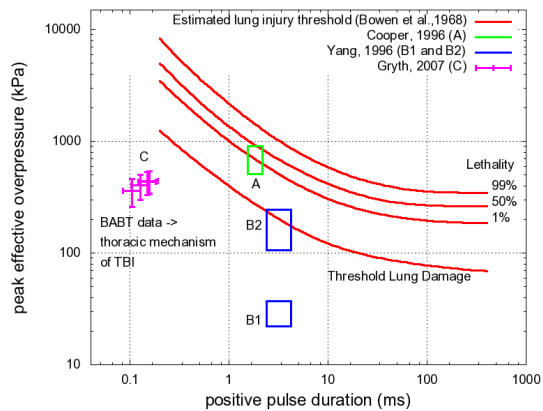

*Figure 2.* Group of data points near C shows behind-armor blunt trauma (BABT) data supporting a thoracic mechanism of blast-induced TBI. Vertical error bars represent the standard error in the regression model used to estimate peak effective pressure from chest wall acceleration. Horizontal error bars represent an estimated 20% uncertainty in the interaction duration. Data are plotted with Bowen curves. EEG suppression, apnea and death were observed following impact of a rifle bullet to the armor-protected thorax of 60 kg swine. (Gryth, 2007, Gryth et al., 2007, Drobin et al., 2007, Gryth et al., 2008).

This analysis of BABT data suggests that the exposure threshold for a thoracic TBI mechanism is below the threshold for lung damage reported by Bowen et al. (1968). Data reported by Gryth et al. (2008) also support a contribution by a vaso-vagal reflex: reduced effects (less EEG suppression, reduced apnea) were observed in vagotomized animals at lower levels of insult. A vaso-vagal component was also observed by Cernak et al. (1996) and Irwin et al. (1999) in animal models exposed to blast waves

Ballistic pressure waves (BPW) initiated by bullet penetration of the thoracic cavity and extremities can also reach the brain and cause neural damage (Suneson et al., 1990a,b; Wang et al., 2004). In 2009, Krajsa reported autopsy results of 33 victims of single, fatal gunshot wounds to the chest (carefully selected to exclude patients with any related history) that consistently showed pericapillar hemorrhage in brain tissue. Courtney and Courtney (2009) discussed the evidence supporting a similar thoracic mechanism of TBI when the pressure waves are due to blast.

Data from BPW studies can be used to estimate an exposure threshold for a thoracic TBI mechanism. These studies include human epidemiologic data and data from human-sized and nearly human-sized animal models (Courtney and Courtney, 2007). In some of these studies, neurotrauma is inferred from reported immediate incapacitation, and this observation is the criterion for inclusion as likely indicating mild TBI. Additional studies report cerebral effects from ballistic pressure waves initiating 0.5m from the brain in a hind limb in pigs or dogs (Suneson et al., 1990a,b; Wang et al., 2004). In these studies, histological and biochemical techniques were used to document remote brain injury, and this evidence of remote brain injury is the criterion for inclusion.

*Table 3:* Representative values from much larger data sets are shown below to illustrate the process of converting mechanical work (W) and penetration depth (d) from BPW studies to peak effective overpressure for comparison with blast studies. The mechanical work (W) is the kinetic energy transferred to the thorax. The peak effective force ($F_{eff}$) is 5W/d, where d has been converted to meters, so the unit of force is newtons. The factor of 5 is an empirical result that finds agreement with measured ballistic pressure waves and represents the fact that the bullet imparts a force approximately 5 times greater on impact with the chest wall than the average force throughout its penetration (Courtney and Courtney, 2007). $M_{eff}$, $a_{eff}$, and $P_{eff}$ are determined the same as in Table 2. The effective interaction time, $t_{eff}$, is an estimate of the interaction time with the chest wall, and is longer than the values in Table 2, because the handgun bullets represented here are slower than the rifle bullets in Table 2. This data is labeled D and E in Figure 3.

| d | W | $F_{eff}$ | $M_{eff}$ | $a_{eff}$ | $P_{eff}$ | $t_{eff}$ | Note |
| mm | J | N | kg | m/s$^2$ | kPa | ms | |
|---|---|---|---|---|---|---|---|
| 250 | 678 | 13560 | 2.0 | 6780 | 384 | 0.251 | i |
| 350 | 339 | 4843 | 2.0 | 2421 | 262 | 0.585 | ii |
| 330 | 89 | 1348 | 2.0 | 674 | 154 | 0.667 | iii |
| 262 | 144 | 2752 | 2.0 | 1376 | 209 | 0.441 | iii |
| 300 | 864 | 14400 | 2.0 | 7200 | 392 | 0.285 | iv |
| 360 | 533 | 7403 | 2.0 | 3701 | 309 | 0.597 | v |

**Notes**: i) From human epidemiological data indicating probable TBI (Courtney and Courtney 2007). ii) From human epidemiological data indicating possible TBI (Courtney and Courtney 2007). iii) From human autopsy data demonstrating pericapillar hemorrhage in brain. (Krajsa, 2009, Cases 12 and 4, respectively). iv) From goat data demonstrating high probability of rapid incapacitation suggestive of likely TBI (Courtney and Courtney 2007). v) From goat data demonstrating moderate probability of rapid incapacitation suggestive of possible TBI (Courtney and Courtney 2007).

Depending on the experimental conditions, various techniques (similar to those used for BABT) were used to estimate a peak blast pressure equivalent to the ballistic impact. Analysis of representative data points from the much larger data sets is described in Table 3. The additional analysis employed for comparison with results of blast studies introduces uncertainty, thereby limiting the precision of the results. Figure 3 shows the results of these analyses.

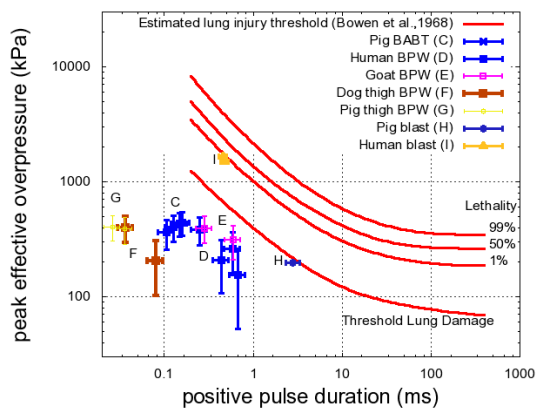

*Figure 3:* Data showing positive indications related to thoracic mechanism of blast TBI. C: BABT data. D: estimates based on analysis of human data of rapid incapacitation or pericapillar hemorrhaging in the brain following a single bullet impact to the thorax (Krajsa, 2009). E: rapid incapacitation was observed following a single bullet impact to the thorax in 70 kg goats (Courtney and Courtney, 2007) F: Ultrastructural damage was observed in the brain following bullet impact to the thighs in 15 kg dogs (Wang et al., 2004). G: Microscopic nerve damage was observed in the brain following bullet impact to the thigh in 20 kg pigs (Suneson et al., 1990b). Points with the higher peak effective overpressure indicate a greater likelihood of mild TBI. Vertical error bars represent the standard error in the regression model used to estimate peak effective pressure from chest wall acceleration. Horizontal error bars represent an estimated 20% uncertainty in the interaction duration. H: EEG suppression was observed in pigs exposed to air blast from behind (Axelsson et al. 2000). I: Immediate neurological effects were reported in humans exposed to underwater blast (Wright et al., 1950).

Analysis of experiments in which test animals were shot in the thigh is more tenuous than that for ballistic impacts to the chest. There is no acceleration of the chest wall analogous to blast exposures and no established effective mass. The mathematical steps can be performed in an analogous manner, but the physical interpretation is less clear. It is possible that the rapid transfer of energy to soft tissues at a high strain rate creates a systemic pressure wave that does not vary greatly between chest and thigh impact. Transmission of the ballistic pressure wave from the thigh to the brain was verified by implanting high speed pressure transducers in the thigh, abdomen, neck, and brain of test animals (Suneson et al., 1990a, 1990b), and unlike the experimental insults to the thorax where the possibility of brain injury was inferred from rapid EEG suppression or observed incapacitation, brain injury was documented via histological and/or biochemical analysis. The soundness of this conversion to peak effective blast overpressure is supported by the direct measurement of abdominal pressures 381±137 kPa (Suneson et al., 1990a, Group III) in the case where our analysis yields 407±102 kPa for peak effective blast overpressure.

*Table 4:* Data shown below illustrate conversion of mechanical work (W) and penetration depth (d) from BPW studies to peak effective overpressure for comparison with blast studies. The mechanical work (W) is the kinetic energy transferred to the thigh tissues. The effective force ($F_{eff}$) is W/d, where d is the tissue depth over which the projectile loses energy, converted to meters. Here, the effective force is estimated as the average force, because the interaction in tissue is much shorter than the total possible penetration, and the peak force is not expected to be much larger than the average in the relatively short depth, in contrast to cases (such as Table 3) where the total penetration is much greater than the region of interest. $M_{eff}$, the effective mass, is estimated to be 1 kg. $a_{eff}$, and $P_{eff}$ are computed the same as in Table 2. This data is labeled F and G in Figure 3.

| d mm | W J | $F_{eff}$ N | $m_{eff}$ kg | $a_{eff}$ m/s² | $P_{eff}$ kPa | time ms | Note |
|---|---|---|---|---|---|---|---|
| 111 | 773 | 6964 | 1.0 | 6964 | 388 | 0.036 | i |
| 79 | 632 | 8000 | 1.0 | 8000 | 407 | 0.026 | ii |
| 100 | 741 | 7409 | 1.0 | 7409 | 396 | 0.037 | iii |
| 100 | 131 | 1312 | 1.0 | 1312 | 204 | 0.080 | iv |

i. Suneson et al., 1990a,b Group I
ii. Suneson et al., 1990a,b Group III
iii. Wang 2004 et al., High Speed Group
iv. Wang 2004 et al., Low Speed Group

Two additional studies informing the thoracic mechanism were also considered. In an experiment by Axelsson et al. (2000), 60 kg pigs were exposed to blast from behind, so that minimal head acceleration occurred, and there was reduced head exposure. Moreover, it is possible that the thicker pig skull resists direct transmission of the blast wave. EEG suppression was observed immediately after the blast. In a study notable for the inclusion of human subjects, Wright et al. (1950) reported immediate neurological effects, including temporary paralysis, when human subjects were exposed to underwater blasts.

*Acceleration mechanisms of primary blast-induced TBI*

In 1980, Ono et al. published head injury tolerance curves based on a combination of primate and human cadaveric test data. In the present study, the acceleration data were adapted to blast parameters via simple dimensional arguments (others are working on detailed FEM). The Ono curve acceleration was converted to force using Newton's second law and a typical head mass of 4.3 kg. The force was converted to pressure using a typical head cross sectional area of 0.035 m$^2$. Other criteria could also be applied, such as the Wayne State head injury criteria (Zhang et al., 2004).

It is uncertain whether rotational acceleration mechanisms are important in primary blast-induced TBI. Translational acceleration may dominate, since blast waves involve much shorter impulses than auto accidents, perhaps too short for gross rotation to begin (Krave et al., 2005).

**Results**

Regions of interest defined by peak effective overpressure and positive pulse duration for blast waves injuring brain tissue via thoracic and acceleration mechanisms are shown in Figures 4 and 5 along with the Bowen thresholds for blast lung injury. While data from the several studies analyzed do not support a definitive threshold, they do support regions of interest to guide experimental designs and compare future data.

For the thoracic mechanism of transmission, the region of interest shown in Figure 4 was drawn to include blast conditions that resulted in neural injury and above blast conditions that did not result in neural injury. The lower limit of the region of interest also excludes exposures that are essentially in the range of loud noises and have been shown to be safe for training purposes (Johnson, 1994; Price, 2005).

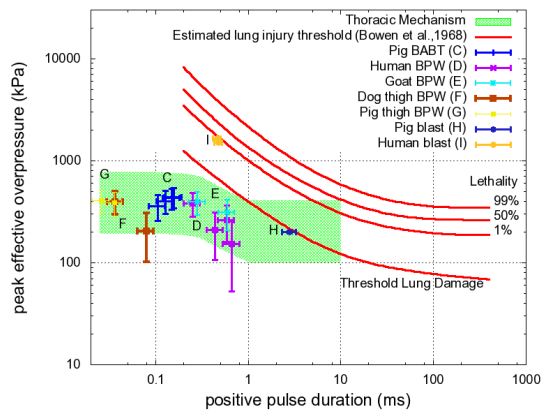

***Figure 4***: Proposed region of interest for blast-induced TBI based on a thoracic mechanism by which pressure waves enter the thorax and lead to brain injury. Bowen curves for blast-induced lung injury are shown for comparison.

The uncertainty associated with the adapted Ono curve in Figure 5 is unknown. The Ono curve was developed using data from impact experiments (Ono et al., 1980), but data from blast experiments isolating the acceleration mechanism are not available. Physical laws and dimensional arguments were used to adapt the curve. Since the uncertainty associated with the adapted curve is unknown, conservative factors of 2 and ½ were chosen to describe a region of interest that is expected to contain future data. However, the lower range of the region of interest for durations of 1 – 10 ms extends into the range deemed safe for training purposes (Johnson, 1994; Price, 2005); therefore, it is not expected that future data will indicate bTBI due to exposures below the adapted Ono curve between 1 and 10 ms positive pulse duration.

Probabilities of injury within each region of interest are not assigned. However, since *a threshold estimates minimum conditions for injury*, increased overpressures at a given duration can be expected to cause more severe injuries.

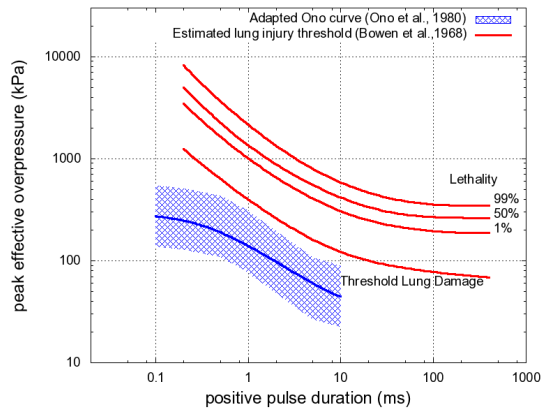

***Figure 5:*** *The Ono curve (adapted via dimensional arguments) and a region of interest defined by factors of 2 and ½ for primary blast-induced TBI due to an acceleration mechanism. The region lies below the Bowen blast-lung thresholds (Ono et al., 1980).*

The adapted Ono curve for TBI due to an acceleration mechanism of blast wave transmission suggests that for a given positive pulse duration, acceleration may lead to TBI at overpressures below those likely to cause lung injury (Figure 5). This curve would be even lower if rotational injury is considered significant at these durations. Increasing confidence that acceleration due to blast waves is contributing to primary blast-induced TBI requires detailed finite element modeling of translational and rotational head and brain accelerations under realistic conditions, as well as blast experiments with a biofidelic human head surrogate.

In certain regions, it appears that more than one mechanism could lead to brain injury. There are combinations of peak effective overpressure and positive pulse duration that are contained in regions of interest for both the thoracic and the acceleration mechanisms of blast-induced TBI. Our results also suggest that under certain blast conditions, one or both of these mechanisms may lead to neural injury while lung injury may be absent. There is a region of positive pulse duration between 1 and 10 ms and peak effective overpressure of 20 – 100 kPa in which the results suggest that an acceleration mechanism may lead to TBI without the thoracic mechanism playing a role and without concomitant lung injury.

At longer durations, multiple contributors to a given mechanical mechanism of blast-induced brain injury may come into play, thereby changing the shape of the proposed exposure thresholds. For example, the threshold for the acceleration mechanism might turn sharply downward as durations become long enough for a rotational injury mechanism (twisting of brainstem) to contribute (Krave et al., 2005). Behavior of the threshold for the thoracic mechanism at longer blast durations depends on the specific nature of the thoracic mechanism, i.e. whether a vascular surge, a high-frequency transmission of energy and/or a vaso-vagal response is leading to injurious conditions in the brain tissue. Despite this uncertainty, the exposure threshold presented here for the thoracic mechanism is consistent with the results of experiments that focused a blast wave on the thorax in an animal model (Cernak et al., 2001a,b).

**Discussion**

Regions of blast overpressure and duration in which thoracic and acceleration mechanisms may result in brain injury were developed based on data from blast studies and from related studies of behind-armor blunt trauma and ballistic pressure waves. Data from the related fields were analyzed to supplement the sparse data available from nonlethal blast studies because the experimental conditions produce blast-like pressure waves and similar injuries. Moreover, the study designs are suited to isolating possible mechanisms of transmission of blast waves to brain tissue and thus can inform the design of new experiments and be compared with additional data as they become available. Additional data are needed to clarify actual probabilities or severity of TBI for a given exposure.

If, as the results suggest, the mechanisms of blast-related TBI are not mutually exclusive, more than one may contribute to an individual's injury. Recall the situation when Louis Pasteur was asked to find the cause of disease devastating the French silkworm industry in 1865 (Dubos, 1950). After a year of diligent work he correctly identified a culprit organism and gave practical advice for developing a healthy population of moths. However, when he tested his own advice, he found disease still present. It turned out he had been correct but incomplete – there were two organisms at work.

It is premature to greatly favor one likely mechanical mechanism of a blast wave leading to TBI to the exclusion of others, lest the mistakes of history be repeated. Well-designed investigations of likely mechanisms should go forward in parallel. The solution to the problem will be expedited if investigators identify – during the design stage if possible – what mechanism(s) are being investigated, how a given experiment or model informs and is informed by likely injury thresholds, and where the work fits in the context of other investigations.